\documentclass[traditabstract]{aa}
\usepackage{txfonts}
\usepackage{graphicx}
\usepackage{natbib}
\bibpunct{(}{)}{;}{a}{}{,} 
\usepackage{epsfig}

\begin{document}
\title{The Rossiter-McLaughlin Effect of\\
CoRoT-3b \& HD\,189733b\thanks{using observations with the \textit{Harps} spectrograph from the ESO 3.6\,m installed at La Silla, Chile, under the allocated programmes 072.C-0488(E) and 079.C-0828(A). The data is publicly available at the \textit{CDS} (\textsc{cdsarc.u-strasbg.fr})}}
\author{Amaury H.M.J. Triaud\inst{1}
\and Didier Queloz\inst{1}
\and Fran\c cois Bouchy\inst{2,3}
\and Claire Moutou\inst{4}
\and Andrew Collier Cameron\inst{5}
\and \\Antonio Claret\inst{6}
\and Pierre Barge\inst{4}
\and Willy Benz\inst{7}
\and Magali Deleuil\inst{4}
\and Tristan Guillot\inst{8}
\and Guillaume H\'ebrard\inst{2}
\and \\Alain Lecavelier des \'Etangs\inst{2}
\and Christophe Lovis\inst{1}
\and Michel Mayor\inst{1}
\and Francesco Pepe\inst{1}
\and St\'ephane Udry\inst{1}
}

\offprints{Amaury.Triaud@unige.ch}

\institute{Observatoire de l'Universit\'e de Gen\`eve, Chemin des Maillettes 51, CH-1290 Sauverny, Switzerland
\and Institut d'Astrophysique de Paris, CNRS UMR 7095, Universit\'e Pierre \& Marie Curie, 98bis boulevard Arago, 75014 Paris, France
\and Observatoire de Haute-Provence, CNRS USR 2207, 04879 St Michel de l'Observatoire, France
\and Laboratoire d'Astrophysique de Marseille, CNRS UMR 6110, Traverse du Siphon, 13376 Marseille, France
\and School of Physics \& Astronomy, University of St Andrews, North Haugh, KY16 9SS, St Andrews, Fife, Scotland, UK
\and Instituto de Astrof\'isica de Andaluc\'ia, CSIC, Apartado 3004, 18080 Granada, Spain 
\and Physikalisches Institut Universit\"at Bern, Sidlerstrasse 5, CH-3012 Bern, Switzerland
\and Observatoire de la C\^ote dÕAzur, Laboratoire Cassiop\'ee, CNRS UMR 6202, BP 4229, 06304 Nice Cedex 4, France}

\date{Received date / accepted date}
\authorrunning{Triaud et al.}
\titlerunning{Rossiter-McLaughlin effect on CoRoT-3b \& HD\,189733b}

\abstract{We present radial-velocity sequences acquired during three transits of the exoplanet HD\,189733b and one transit of CoRoT-3b. We applied a combined Markov-chain Monte-Carlo analysis of spectroscopic and photometric data on these stars, to determine a full set of system parameters including the projected spin-orbit misalignment angle of HD\,189733b to an unprecedented precision via the Rossiter-McLaughlin effect: $\beta = 0.85 ^{\circ\,+ 0.32}_{\,\,\,\,- 0.28}$. This small but non-zero inclination of the planetary orbit is important to understand the origin of the system. On CoRoT-3b, results seem to point towards a non-zero inclination as well with $\beta=37.6^{\circ\,+ 10.0}_{\,\,\,\,- 22.3}$, but this remains marginal. Systematic effects due to non-gaussian cross-correlation functions appear to be the main cause of significant residuals that prevent an accurate determination of the projected stellar rotation velocity $V\,sin(I)$ for both stars.

 }

\maketitle

\section{Introduction}

The spectroscopic transit, also known as the Rossiter-McLaughlin (RM) effect (\citet{Rossiter:1924p869, McLaughlin:1924p872}), is a radial velocity (RV) anomaly superimposed on the radial-velocity curve arising from the Keplerian reflex orbit of the host star about its common centre of mass with a planet.
As the planet transits, it covers - in the case of a prograde orbit - first the blue shifted part of the rotating star, shifting the overall spectrum slightly to the red. As the planet moves across the stellar disc, the radial velocity of the star's light centroid changes rapidly. The first RM effect caused by a transiting planet was observed on HD\,209458 by \citet{Queloz:2000p247}. In addition to the standard information that a photometric transit brings us, the RM effect permits us to measure the $V\,sin(I)$\footnote{We used the \citet{Ohta:2005p631} notation differentiating the $V\,sin(i)$ to the $V\,sin(I)$. $i$ is the projected inclination of the planet's orbit on the sky, whereas $I$ is the projected inclination of the stellar equator on the sky.} of the star and is the only way to estimate the projected spin-orbit obliquity angle $\beta$  \citep{Gimenez:2006p31, Hosokawa:1953p2002} (equal to $-\lambda$ \citep{Ohta:2005p631}). This angle reflects the history of the planet and can therefore be used to constrain models of planetary orbital evolution.

HD\,189733 is a most interesting target due to its brightness and the possibility of observing it from both hemispheres fairly easily, while CoRoT-3b, orbiting a very fast rotating star and being itself in the middle of the Brown Dwarf desert \citep{Deleuil:2008p408} (D08), is a unique object to study. Also, HD\,189733b has been extensively studied (\citet{Bouchy:2005p828}, \citet{Winn:2006p93} (W06), \citet{Winn:2007p440} (W07), \citet{Boisse:2009p1077} (B09)). The present paper is motivated by our recent acquisition of high-precision data - photometric for CoRoT-3, photometric and spectroscopic for HD\,189733. Combining spectroscopic and photometric transits in one analysis allows us to refine the transit parameters, but also to ensure that the $V\,sin(I)$ of the star and the spin-orbit angle $\beta$, which are solely extracted from the RM effect, are tightly constrained. This is especially effective when analysing high precision data ; thus we use the high precision obtained on HD\,189733 and present this star as a test-case for combining data sets.

We describe the observations of these stars in section~\ref{sec:obs}, then move to a description of the fitting process and its results with section~\ref{sec:fit} and explore a few reasons to explain the observed residuals that we obtained in section~\ref{sec:res}. Finally we will discuss the results and conclude.

\section{The Observations}\label{sec:obs}

The observations were acquired using the High Resolution \'echelle spectrograph \textit{Harps} mounted on the 3.6\,m at the ESO observatory of La Silla, in Chile. The data were extracted using the Data Reduction Software present at the telescope. A reanalysis of the data was performed later in Geneva, using the latest version of the software as in \citet{Mayor:2009p1088}. 

One sequence of 11 \textit{Harps} RV measurements in addition to those in the discovery paper (D08) was obtained on CoRoT-3 on August $26^{th}$ 2008 as part of the spectroscopic follow-up (072.C-0488(E)); all are around the transit. On average the new data has an estimated photon noise of $28.4\,m.s^{-1}$. 

Four sequences were taken of HD\,189733 on July $30^{th}$, August $4^{th}$ and September $8^{th}$ 2006, and on August $29^{th}$ 2007 under the allocated programme 079.C-0828(A) and as part of the GTO, three of which are during transit. One sequence was taken off-transit to act as a comparison sequence. Out of the three RM sequences, two were obtained using a low cadence (one point every 10.5 minutes). One of these suffered from bad weather and one was taken with a high cadence (one point every 5.5 minutes). In total we have 78 new RV measurements including 37 during transit. The mean estimated uncertainty in the radial velocities due to photon noise is $0.98\,m.s^{-1}$.

\section{Fitting the Data}\label{sec:fit}
Transiting planets have an important role to play in constraining planetary evolution models as well as atmosphere and interior models, therefore it is important that everything is done to ensure that information extracted from the data is accurate. Both spectroscopic and photometric effects can be observed on a star experiencing a planetary transit. The two types of observation constrain parameters differently but arise from the same cause. Hence it is logical to fit both types of data simultaneously to determine a single set of parameters for the planet and ensure full consistency between the models. 

\subsection{the modelling}\label{subsec:modelling}

A code was developed using full Bayesian statistics in a Markov Chain Monte-Carlo (MCMC). The code is similar to the one described in Collier \citet{CollierCameron:2007p704} and has already been used \citep{Gillon:2008p767, Bouchy:2008p229}. The philosophy here is to combine everything that is known about each star into the fitting process to better constrain the final result. So far, the periodic Doppler shift caused by the planet, the RM effect and a drift can be fitted to the spectroscopy. For the photometry, requiring data stripped of instrumental or stellar effects, primary and secondary transits can be fitted. Limb darkening coefficients can also be allowed to float if data of very high photometric precision are available.

The present version of the code fits up to 12 free parameters: the depth of the primary transit $D$, the RV semi-amplitude $K$, the impact parameter $b$, the transit width $W$, the period $P$, the middle of transit $T_0$, $e\,cos(\omega_0)$ \& $e\,sin(\omega_0)$ (with $e$ being the eccentricity and $\omega_0$ the angle between the line of sight and the periastron), $V\,sin(I)\,cos(\beta)$ \& $V\,sin(I)\,sin(\beta)$, the RV drift $\Gamma$ and the secondary transit depth $D_2$. If some parameters are irrelevant, they can be fixed. These parameters have been chosen to reduce correlations and increase the exploration of parameter space. 

\begin{figure}
\centering                     
\includegraphics[width=8cm]{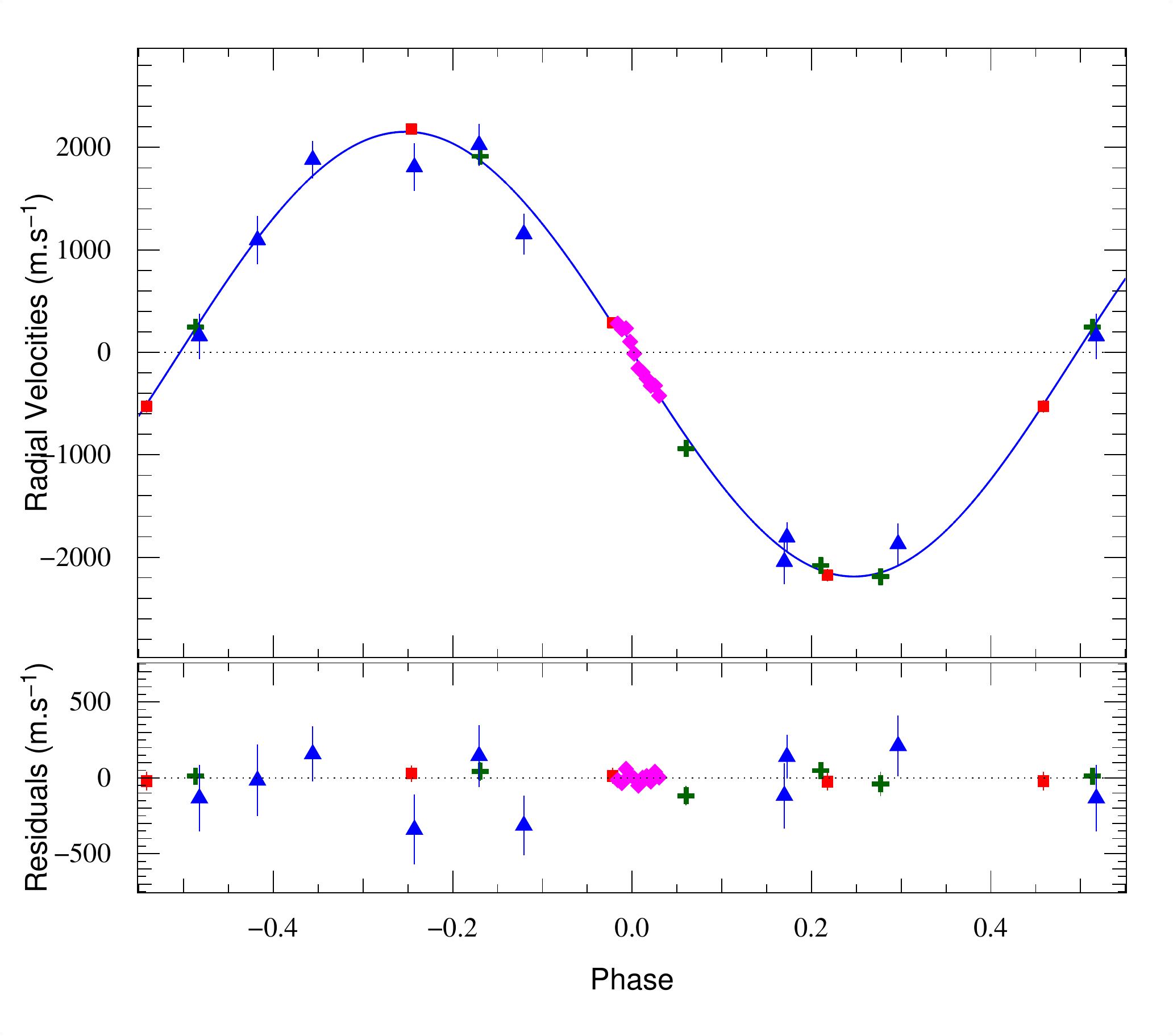}
\includegraphics[width=8cm]{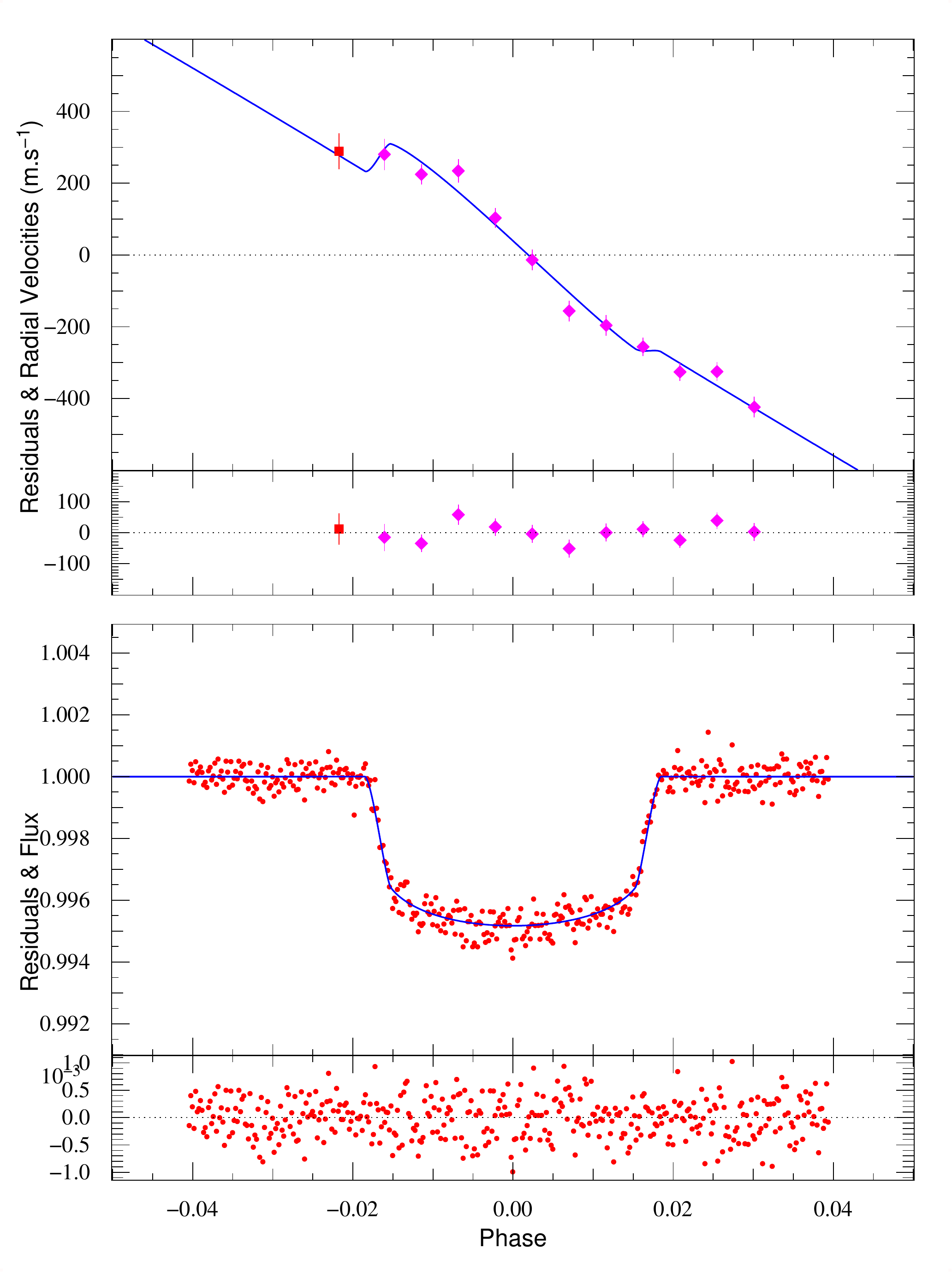}
\caption{{\bf{top}}: Overall Keplerian fit of the RV data for CoRoT-3b.
{\bf{bottom}}: Composite plot showing both the spectroscopic and the photometric transit for CoRoT-3b.
\textit{(red) squares} are \textit{Harps} measurement which are not part of the RM sequence, \textit{(green) crosses} are \textit{Sophie} measurements, \textit{(blue) triangles} show \textit{TLS} data and \textit{(magenta) diamonds} indicate the RM \textit{Harps} sequence; \textit{(red) circles} are for the CoRoT photometry.}\label{fig:corot3}

\end{figure}

At each step $i$ of the Markov chain, one set of $j$ parameters is calculated from the previously accepted value ($i-1$) following:
\begin{equation}\label{eq:a}
P_{i ,j}= P_{i-1,j} + f\,\sigma_{P_j}\,G(0,1)
\end{equation}
where $P_j$ is a parameter, $\sigma_{P_j}$ is the $1\,\sigma$ uncertainty, $f$ is a factor ensuring 25\% of steps are being accepted (see \citet{Tegmark:2004p793}) and $G(0,1)$ is a random Gaussian number centred on zero with a standard deviation equal to 1. From these parameters a large variety of other - more physical - values can be inferred such as the stellar density $\rho_{\star}$ (see Table~\ref{tab:params}). 

\begin{figure}
\centering                     
\includegraphics[width=8cm]{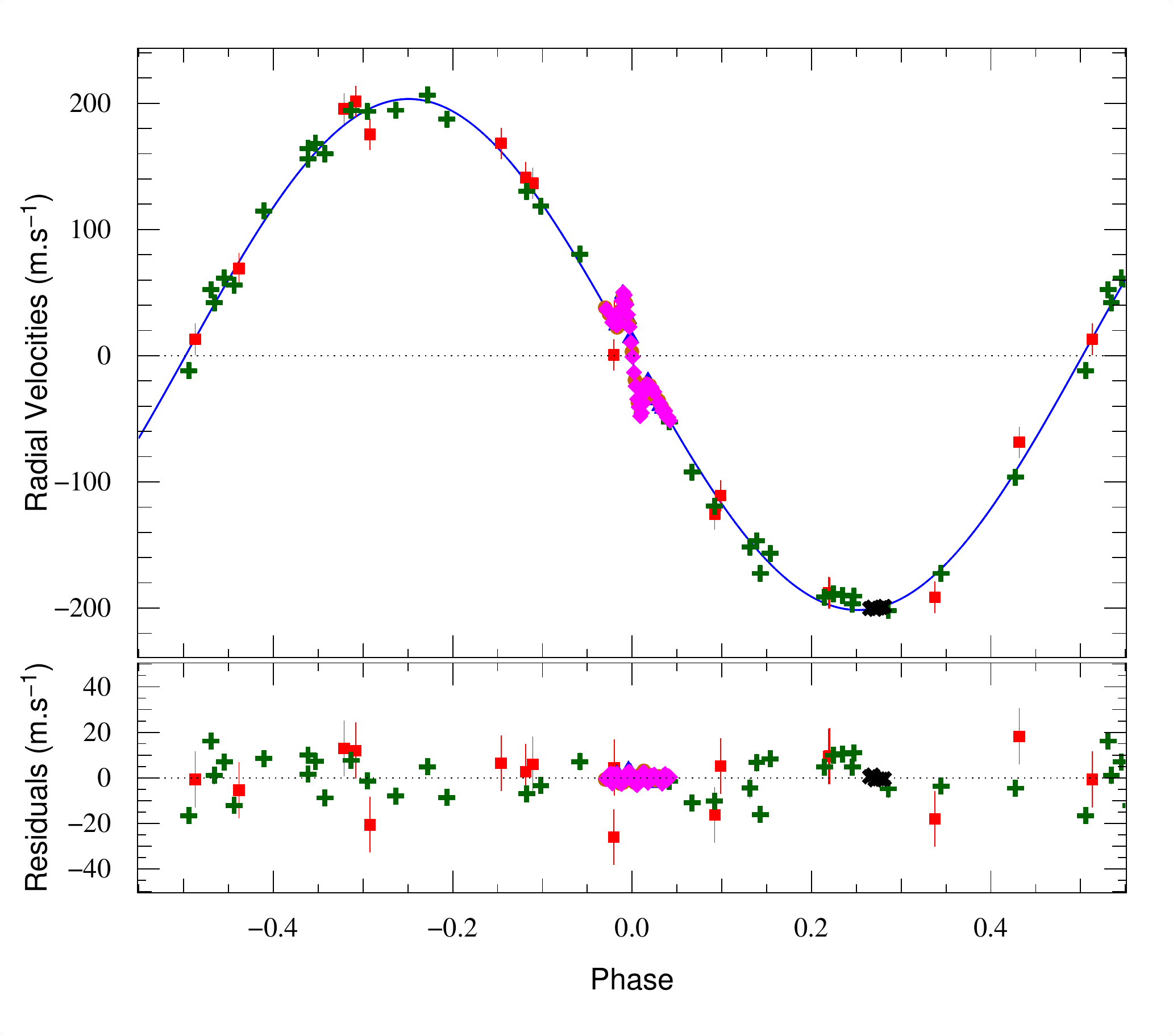}
\includegraphics[width=8cm]{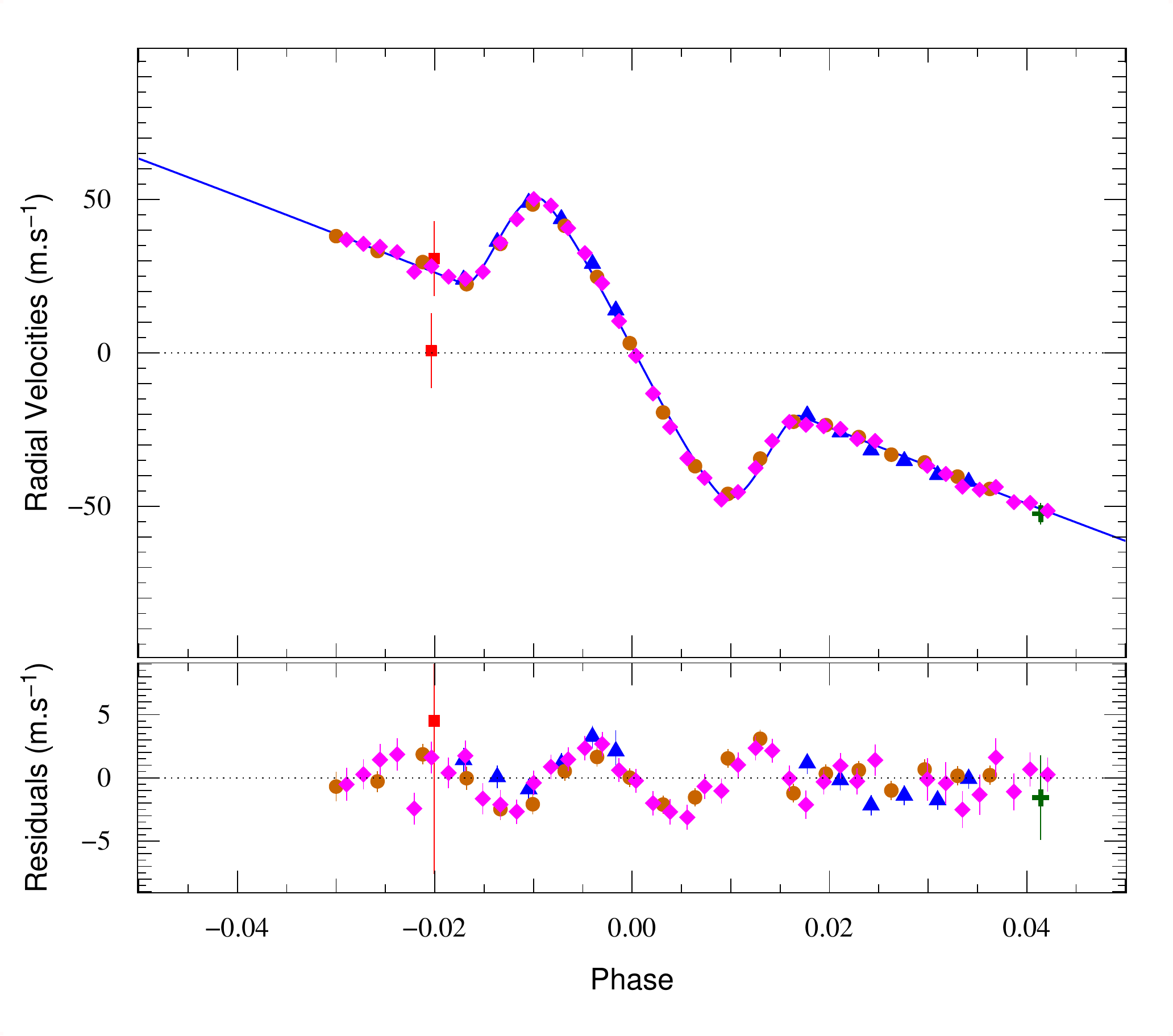}
\caption{{\bf{top}}: Overall Keplerian fit of the RV data for HD\,189733b.
{\bf{bottom}}: RM effect on HD\,189733. \textit{(red) squares} shows \textit{Keck} data from W06, \textit{(green) crosses} are \textit{Sophie} data from B09, \textit{(blue) triangles}, \textit{(orange) circles}, \textit{(magenta) diamonds} are the new three \textit{Harps} sequences on the RM effect and \textit{(black) oblique crosses} show the off transit \textit{Harps} sequence. Residuals are displayed below and show a clear correlated signal during the transit.}\label{fig:HD189}

\end{figure}

The parameters are then used to calculate three different models. Both primary and secondary transit are calculated by using either codes developed by \citet{Mandel:2002p768}, or by \citet{Gimenez:2006p40}. The RV curve is modelled by the standard orbital equations \citep{1995Natur.378..355M, Hilditch:2001p773} and by the code presented in \citet{Gimenez:2006p31} for the RM effect. To model the stellar limb darkening, we use the quadratic law \citep{Claret:2000p364}. The model and the data are compared using a $\chi^2$ statistics. 

$\chi^2$ from the photometry is added to the value found for the spectroscopy. On that value Bayesian penalties are added. These can be estimated for every parameter for which we have independent prior knowledge of their value and error.  The stellar mass $M_{\star}$ is allowed to vary subject to a Bayesian penalty on a value and its $1\,\sigma$ error bar. All added, it creates a merit function:

\begin{equation}\label{eq:b}
Q_i = \chi^2_i + {(M_{\star_i}-M_{\star_0})^2\over{\sigma_{M_{\star}}^2}} + {\sum^P_{j=1}{{(P_{i,j}-P_{0,j})^2\over{\sigma_{P_j}^2}} }}
\end{equation}
where here, $P_j$ can be any parameter, fitted or physical and $P_{0,j}$ is the value of the prior as $M_{\star}$ is the floating parameter for stellar mass and $M_{\star_0}$ is the prior on the mass.

This $Q_i$ is compared to the $Q_{i-1}$, value calculated from the previous set of parameters, with the Metropolis-Hasting algorithm. This is repeated as many times as is necessary to ensure that the fitting has converged and that the exploration of parameter space around the best value is truly random (meaning that the correlation length for each parameter is small compared to the number of accepted steps) and gives credible error bars.

Obviously in addition to all these parameters, we also have to add one $\gamma$ velocity for each RV set and one normalisation constant for each photometric set; they are estimated by optimal averaging and optimal scaling in $\chi^2$ calculations. The $\gamma$ velocity reflects the mean radial velocity due to the motion of the star in comparison to the Sun; its measured value varies with activity levels and between instruments by a small offset. The normalisation factors re-estimate the normalisation of the lightcurves. The best fit parameters are chosen to be the set with the lowest $\chi^2$ and the error bars are calculated by taking the 68.3\% lowest values of $\chi^2$ and finding their extremes.

\begin{table}
\caption{Fitted and physical parameters found after fitting photometric and spectroscopic data of CoRoT-3b and of HD\,189733b. The reduced $\chi^2$ were calculated with 414 degrees of freedom for CoRoT-3b and 2890 for HD189773b. \textbf{Nota Bene} $\beta = -\lambda$; $\beta$ is used since the first reference of a projected spin-orbit angle was named thus, in \citet{Hosokawa:1953p2002}. $V\,sin(I)$s are probably overestimated, see section~\ref{sec:res} for details.}\label{tab:params}
\begin{tabular}{lll}
\hline
\hline
Parameters $(units)$ & CoRoT-3b & HD\,189733b \\
\hline
&&\\
\textit{fitted parameters} &&\\
&&\\
$D$                              &$ 0.004398 ^{+ 0.000084}_{- 0.000091}$         & $ 0.0200^{+ 0.00015}_{- 0.00017} $   \\
$K$ $(m.s^{-1})$       &$ 2169.9 ^{+ 35.1}_{- 22.7}$                                 & $ 201.96^{+ 1.07 }_{- 0.63 } $  \\
$b$ $(R_{\star})$       &$ 0.54 ^{+ 0.041}_{- 0.081}$                                & $ 0.6873 ^{+ 0.0047 }_{- 0.0078 } $ \\
$W$ (days)                 &$ 0.1566 ^{+ 0.0012}_{- 0.0014 }$                      & $ 0.07527^{+ 0.00020}_{- 0.00037} $  \\
$P$ (days)                  & $ 4.2567994 ^{+ 0.0000039}_{- 0.0000031}$ & $2.21857312^{+ 0.00000036}_{- 0.00000076} $\\
$T_0$ (bjd)                 &$ 54283.13388 ^{+ 0.00026}_{- 0.00022}$         &$53988.80339^{+ 0.000072}_{- 0.000039} $\\
$e\,cos(\omega_0)$  &$ -0.0083 ^{+ 0.0054}_{- 0.0041}$                     & $0.0038^{+0.0020}_{-0.0020}$\\
$e\,sin(\omega_0)$   &$ 0.000 ^{+ 0.021}_{- 0.020}$                             & $-0.0017^{+0.0024}_{-0.0034}$\\
$V\,sin(I)\,cos(\beta)$ &$ 28.4 ^{+ 6.5}_{- 5.6}$                                        & $3.316^{+0.017}_{-0.068}$ \\
$V\,sin(I)\,sin(\beta)$  &$ 21.9 ^{+ 8.3}_{- 14.1}$                                      & $0.049^{+0.018}_{-0.017}$ \\
$\Gamma$ $(m.s^{-1}.yr^{-1})$           & -                                                   &  $ -0.2 ^{+ 2.7}_{- 3.9}$\\
      &      &  \\
      &      &  \\
\textit{derived parameters}       &      &  \\
&&\\
$R_p / R_{\star}$                         & $ 0.06632 ^{+ 0.00063}_{- 0.00069} $  & $ 0.15812 ^{+ 0.00046}_{- 0.00052}$ \\
& & \\
$R_{\star} / a$                               & $ 0.1257^{+ 0.0057 }_{- 0.0064} $  & $ 0.1142 ^{+ 0.0006}_{- 0.0012} $\\
$\rho_{\star}$   $(\rho_{\odot})$ & $ 0.372^{+ 0.064 }_{- 0.047} $         &$ 1.831 ^{+ 0.059}_{- 0.029}$  \\
$R_{\star}$  $(R_{\odot})$          & $1.540^{0.083}_{0.078}$                  & $ 0.766 ^{+ 0.007}_{- 0.013}  $\\
$M_{\star} $  $(M_{\odot})$        & $ 1.359^{+ 0.059 }_{- 0.043}$          & $ 0.823 ^{+ 0.022}_{- 0.029} $\\
$V\,sin(I)$ $(km.s^{-1})$              &$ 35.8 ^{+ 8.2}_{- 8.3}$                      &$ 3.316 ^{+ 0.017}_{- 0.067}$ \\
& & \\
$R_p / a$                               & $ 0.00834^{+ 0.00042 }_{- 0.00050} $  & $ 0.01805 ^{+ 0.00011}_{- 0.00025} $\\
$R_p$  $(R_J)$                    & $ 0.9934^{+ 0.058}_{- 0.058} $              & $ 1.178 ^{+ 0.016}_{- 0.023} $\\
$M_p$ $(M_J)$                    & $21.23 ^{+ 0.82}_{- 0.59}$                       & $ 1.138 ^{+ 0.022}_{- 0.025} $\\
&&     \\
$a$ $(AU)$                        & $ 0.05694 ^{+ 0.00096}_{- 0.00079} $      &$ 0.03120 ^{+ 0.00027}_{- 0.00037} $\\
$i$   $(^{\circ})$                 & $ 86.10 ^{+ 0.73}_{- 0.52} $                        & $ 85.508 ^{+ 0.10}_{- 0.05} $\\
$e$                                      & $  0.008 ^{+ 0.015}_{- 0.005} $                   & $ 0.0041 ^{+ 0.0025}_{- 0.0020} $\\
$\omega_0$  $(^{\circ})$ & $179 ^{+ 170}_{- 170}$                               &  $ -24.1 ^{+ 33.9}_{- 34.5}$\\
$\beta$ $(^{\circ})$           & $37.6^{+10.0}_{-22.3}$                               &  $ 0.85 _{- 0.32}^{+ 0.28}$\\
\\
\\
$\gamma$ \textit{velocities} ($m.s^{-1}$)\\
\\
\textit{Sophie}&-56182.46& -2273.59\\
\textit{TLS}&-56652.08&\\
\textit{Keck}&-&-15.84\\
\textit{Harps}&-56156.08&-2161.14\\
\textit{Harps} (RM)&-56160.84&-2191.92\\
&-&-2225.44\\
&-&-2204.07\\
\\
\textit{normalisation factors}\\
\\
\textit{CoRoT} & 0.9999998&-\\
\textit{FLWO 1.2\,m z-band} &-&0.99974 \\
\textit{T10 (b+y)/2 band} &-&0.99977 \\
\textit{MAGNUM 2\,m V-band} &-&0.99979 \\
\textit{Wise 1\,m I-band} &-&0.99977 \\
\\
\\
$\chi^2_{reduced}$           & $1.17\pm0.08$ &  $ 1.21\pm 0.03$\\
&&\\
\hline

\end{tabular}

\end{table}

\subsection{the results}\label{subsec:results}

One chain of 500,000 accepted steps was calculated for each star.
Results for both stars appear in Table~\ref{tab:params}. The corresponding fits are displayed in Fig.~\ref{fig:corot3} and Fig.~\ref{fig:HD189}. 

\subsubsection{CoRoT-3b}\label{Corot3}

Nine free parameters, four $\gamma$ velocities and one photometric normalisation factor were used for CoRoT-3b to fit four RV sequences totalling 29 measurements and one sequence of photometry - the binned and phase folded data from D08 - with 400 points in it. This amounts to 414 degrees of freedom. Results between this last paper and the present analysis are not very different. The transit spectroscopic sequence covers little outside the RV anomaly of the RM effect and due to the faintness of the star and poor sampling during the transit, $V\,sin(I)$ and $\beta$ are not well defined. $V\,sin(I)$ is found abnormally large at $35\pm8\,km.s^{-1}$  (see sect.~\ref{subsec:model}); $\beta$ is different from zero only at the $2\,\sigma$ level. 

Bayesian penalties were imposed for priors on the stellar mass $M_{\star_0} = 1.37 \pm 0.09\, M_{\odot}$ and also for the period $P_0 = 4.25680\pm0.000005$ days because we made use of an already folded lightcurve as the data from D08 was released, rather than individual transits; the RM effect was not strong enough to constrain $P$. $T_0$ was allowed to float and is found to differ from the published value of $54283.1383\pm 0.0003$. This is possibly an artefact due to using a folded lightcurve. The RM sequence also suffers from a lack of \textit{continuum} on either side of the transit. We employed the same limb darkening coefficients as D08.

Our initial fit had a Bayesian penalty on the $V\,sin(I)$ for the value of $17\pm1\,km.s^{-1}$ published in D08. Removing the penalty allowed us to find the current value of $35\pm8\,km.s^{-1}$ and permitted a minimisation of $\chi^2$ on the spectroscopy. See table~\ref{tab:corot3ros} where we show that having a $V\,sin(I) = 35\,km.s^{-1}$ and a $\beta = 37.6^{\circ}$ is a significant improvement of the model compared to an aligned system with the spectral analysis value of  $V\,sin(I) = 17\,km.s^{-1}$ (because of a difference in $P$ and $T_0$ we also show results using parameters found by D08).

We have 11 RV data points and three free parameters here (one $\gamma$ velocity, $V\,sin(I)$ and $\beta$) making the total number of degrees of freedom eight. Thanks to the photometry, we have a secured detection of the planetary transit and of the Keplerian orbit and hence have a 100\% chance that the RM effect will occur. It is only a matter of $V\,sin(I)$ being different from zero. We obtain a $\chi^2_{reduced} = 1.575$ just for the RM effect which has to be compared with a $\chi^2_{reduced} = 5.962$ if we adjust with a $V\,sin(I) = 0$. Hence, there is a clear detection of an RV deviation from the Keplerian orbit at the location of the RM effect.  

\begin{table}
\caption{Comparing $\chi^2$ for various solutions proposed for CoRoT-3b.}\label{tab:corot3ros}
\begin{tabular}{cccc}
\hline
\hline
$\chi^2$ & $P$, $T_0$ & $V\,sin(I)\,(km.s^{-1})$ & $\beta\, (^{\circ})$ \\
\hline
\\
$12.6\pm5.0$ & this paper & 35 & 37.6\\
$24.1\pm6.9$	& this paper & 17 & 0\\
$47.7\pm9.8$ & this paper & 0 & 0 \\
\\
$14.0\pm5.3$ & D08 & 35 & 37.6 \\
$23.2\pm6.8$ & D08 & 17 & 0 \\
$47.7\pm9.8$ & D08 & 0 & 0 \\
\\
\hline
\end{tabular}

\end{table}

\subsubsection{HD\,189733b}\label{HD189733}

A total of 127 RV measurements were used for the fit including those published in B09 \& W06 outside of the RM region. The RM effect in W06 is not used here. The RV data were fitted along with 2735 photometric points representing data from four transits in the $z$, $(b+y)/2$, $V$ and $I$ bands from W07. Compared to CoRoT-3b, 11 parameters were allowed to float, adding a drift $\Gamma$ and $P$, plus two $\gamma$ velocities and three photometric normalisation factors more. This gives 2890 degrees of freedom.

We applied a Bayesian penalty on $M_{\star}$ alone, with the value $0.82\pm0.03\,M_{\odot}$ \citep{Bouchy:2005p828}, i.e. using a $Q_i$ from eq.~\ref{eq:b} with the two first terms only. We obtain very good constraints on $P$ \& $T_0$ to a similar order of magnitude compared to \citet{Agol:2008p476} who used five \textit{Spitzer} transits and four secondary transits. Most notable is the precision on the spin-orbit angle $\beta = 0.85 ^{\circ\,+ 0.32}_{\,\,\,\,- 0.28}$, a 99.92\,\% ($3\,\sigma$) confident detection (W06 had found $\beta = 1.4^{\circ}\pm1.1$). This value is found by comparing $\chi^2$ with the value obtained by fixing $\beta=0$ and calculating the probability that indeed $\chi^2$ was improved.

HD\,189733 is known to be active (\citet{Bouchy:2005p828}), and stellar activity (stellar spots) causes changes in RV. Because the activity is acting on a timescale of the order of the rotational period, each of the four \textit{Harps} sequences are expected to have no jitter related to stellar activity. But, the sequences were taken at different epochs, therefore at different activity levels. This is why each of the four sequences is fitted with its own $\gamma$ velocity. The change in $\gamma$ velocity of $33.5\,m.s^{-1}$ is consistent with activity levels found by W06 and by B09.

The results on HD\,189733 confirm it is an exceptional target against which to test the models, fitting techniques and data extraction.

\section{Residual analysis}\label{sec:res}

It is clear when looking at the residuals of HD\,189733b's RM effect that a systematic error is present.  The \textit{RMS} of points within transit is 57\,\% larger than outside and their dispersion is correlated the same way for all three sequences. We also note that the residuals have a form very similar to those found by W06 (see Fig. 1 of that paper).

On CoRoT-3 the \textit{RMS} is comparable to the mean error bar outside and inside the transit indicating a good fit of the RV anomaly caused by the RM effect. 

For both stars we find $V\,sin(I)$ values larger than those present in the literature and found by an analysis of the spectral lines. For HD\,189733, we have a 10\,\% difference (or $2\,\sigma$ from the accepted value); for CoRoT-3, the value found is twice that which is inferred from spectral line analysis.
We considered several possible causes of the effect observed in the residuals of HD\,189733b and the large $V\,sin(I)$ found on both stars.

\subsection{limb darkening}\label{subsec:limb}

We assumed that \textit{Harps} was centred on the $V$ band and limb darkening coefficients were chosen accordingly. The \textit{Harps} spectral response was used to determine a new table of coefficients in the manner described in \citet{Claret:2000p364}. The difference between those newly found coefficients was small: $u_a = 0.6454$ instead of $0.6355$ and $u_b =  0.1375 $ instead of $0.1488 $. $\chi^2$, for just the three HD\,189773b RM \textit{Harps} sequences, passed  from 213 to 206 $\pm\,20$, consistent with no change at all. The shape in the residuals was not altered; this difference in limb darkening coefficients cannot explain the problem.

Limb darkening being less well constrained near the limbs than at centre of the star, we omitted observations taken during ingress and egress, and re-fitted the RM effect. This tested whether the limb-darkening law itself could be causing the problem. The fit was no better, the shape in the residuals was still present, ruling out this possibility as well.

\subsection{differential rotation}\label{subsec:rotation}

\begin{figure}
\centering                     
\includegraphics[width=8cm]{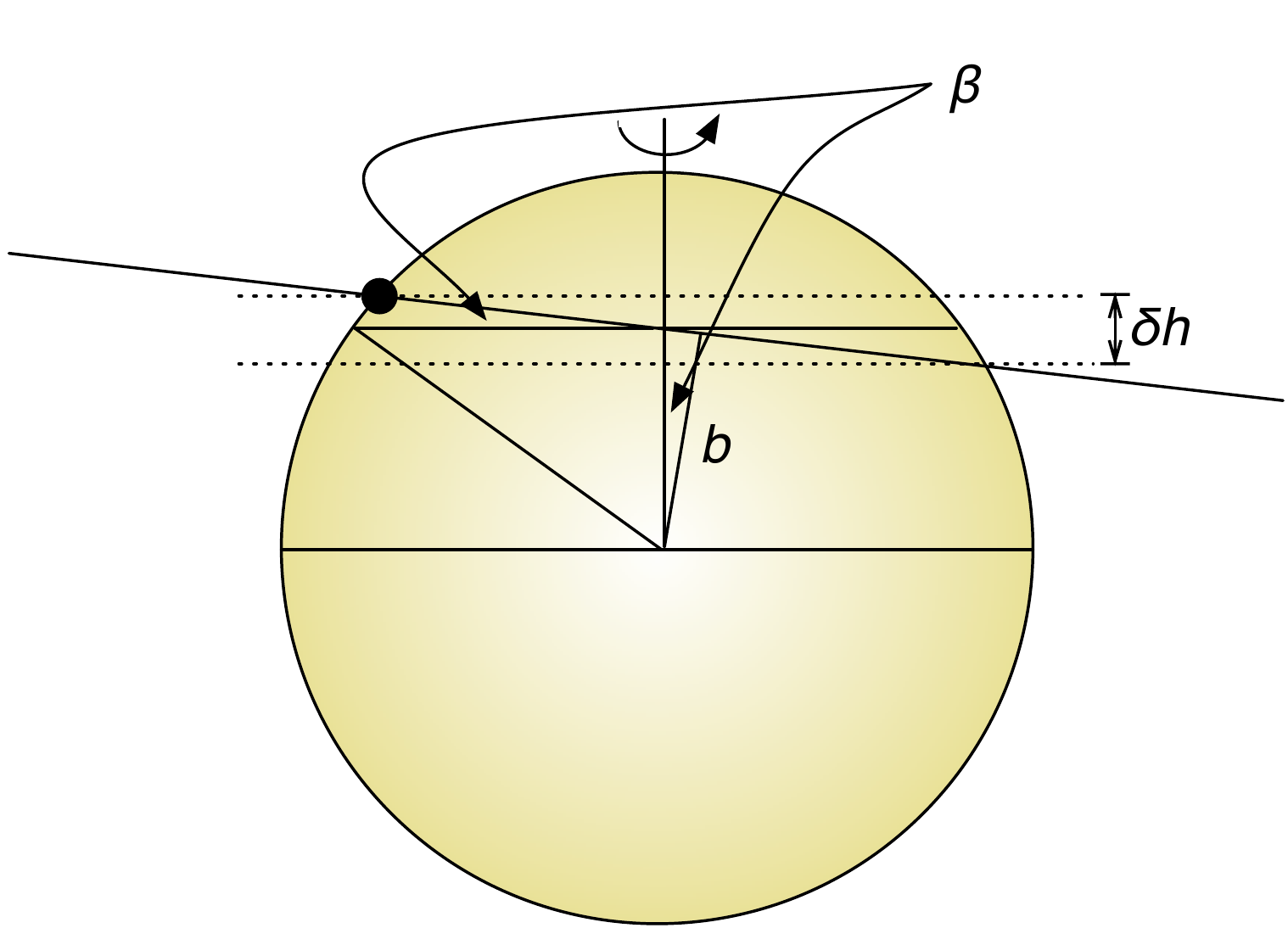}
\caption{Problem geometry and parameters used in equations present in section~\ref{subsec:rotation}}

\label{fig:c}

\end{figure}

The Sun exhibits surface differential rotation as a function of latitude. It is reasonable to assume that other solar like stars rotate differentially. The change in apparent rotational velocity for HD\,189733b, thanks to a misaligned orbital angle and by taking a value of differential rotation to be 10\,\% between the stellar poles and the equator, affects the amplitude of the RM effect very little:

\begin{equation}
{\Delta\,V\,sin(I) } = 0.3\,km.s^{-1}.\,\Delta\,R_{\star}
\end{equation}

\begin{equation}
\delta\,h \simeq |\,2\,.  \sqrt{1 - (b\,.\,cos\,\,\beta)^2}\,.\,tan\, \beta\,\,| = 0.0218\,R_{\star}
\end{equation}
\begin{equation}
{\delta\,V\,sin(I)}= {\Delta\,V\,sin(I) }\,.\,\delta\,h = 0.0065\,km.s^{-1}
\end{equation}
with $h$ the altitude of the planet's path above the stellar equator, projected onto the star (see Fig.~\ref{fig:c}). The value is an order of magnitude lower than our error bars.

It would also be expected to have an effect similar to a non zero spin-orbit angle: a difference in the amplitude on either side. Here the residuals are symmetric with respect to the centre of the transit. This said, with the level of precision now obtained on the spin-orbit angle of HD\,189733b, it would be interesting to see how much of the $\beta$ value is due to differential rotation, but this analysis goes beyond the scope of this paper. It can be emphasised here that thanks to misaligned planets and in conjonction with more precise observations the Rossiter-McLaughlin effect will allow to detect differential rotation on stars without needing the spots that are used at the moment by Doppler tomography on stars like AB Dor.

\subsection{a systematic effect in the data reduction}\label{subsec:model}

\citet{Winn:2005p83} (W05) pointed out that by fitting a symmetric Gaussian to a - by definition - non symmetric and varying cross correlation function (CCF) of the spectra in order to find the RVs and comparing these with a model which takes the centroid of the velocity weighted by the light emitted, we introduce a systematic error. The RV amplitude based on the symmetric Gaussian fit would be larger than that which a model with the same parameters would create. W05 tried to correct this effect by adding a polynomial on the RM model. This action was repeated on W06 and subsequent papers. 

In W05, it was not shown what effect such a misuse of the model would have on the fit. Thus, we created two models: one would act as the theoretical RM model does (Ohta's or Gimenez's), the other recreates the way the spectroscopic data is affected by the transiting planet, and the way the RVs composing the RM sequences are extracted.

\begin{figure}
\centering                     
\includegraphics[width=8cm]{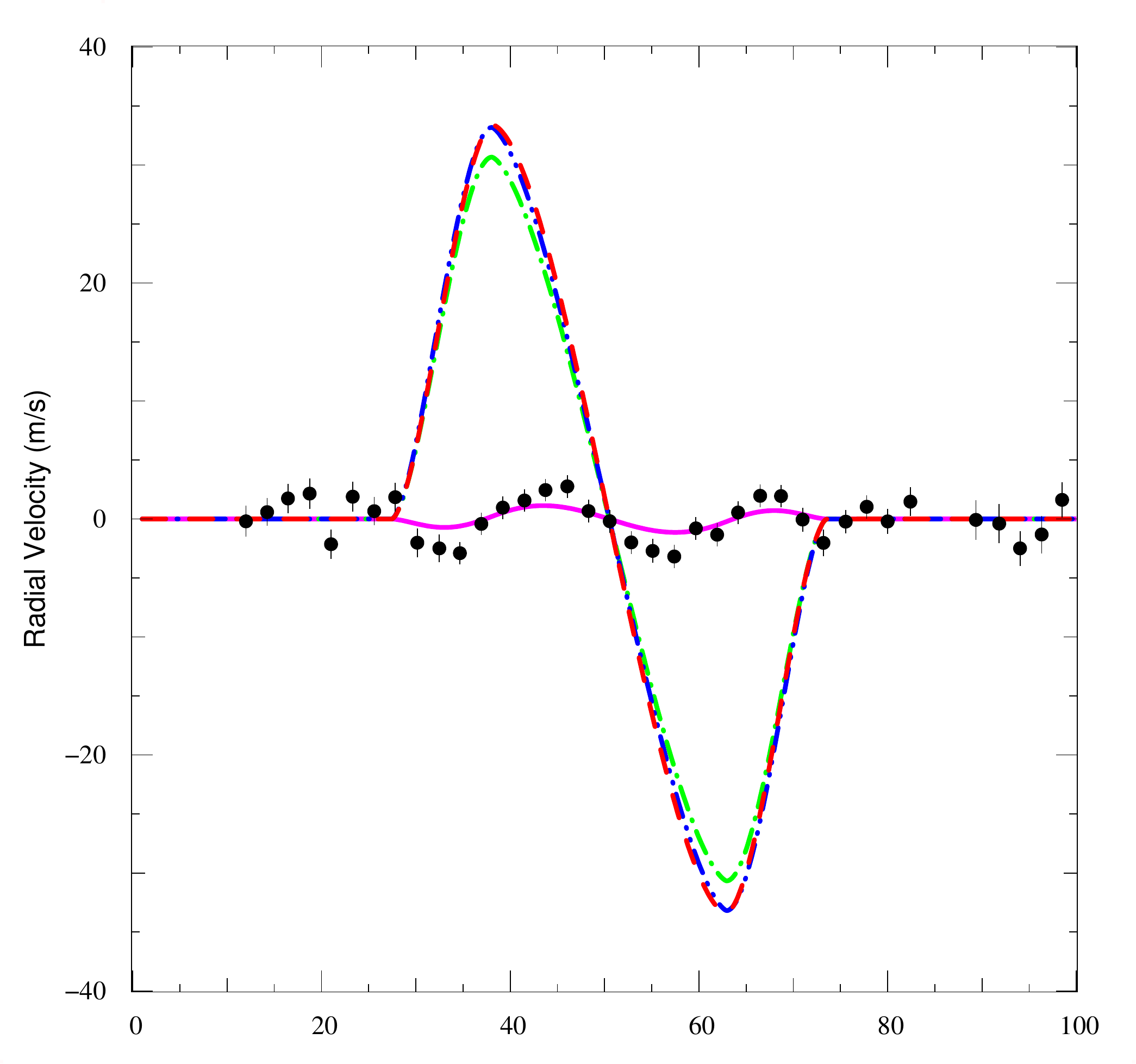}
\includegraphics[width=8cm]{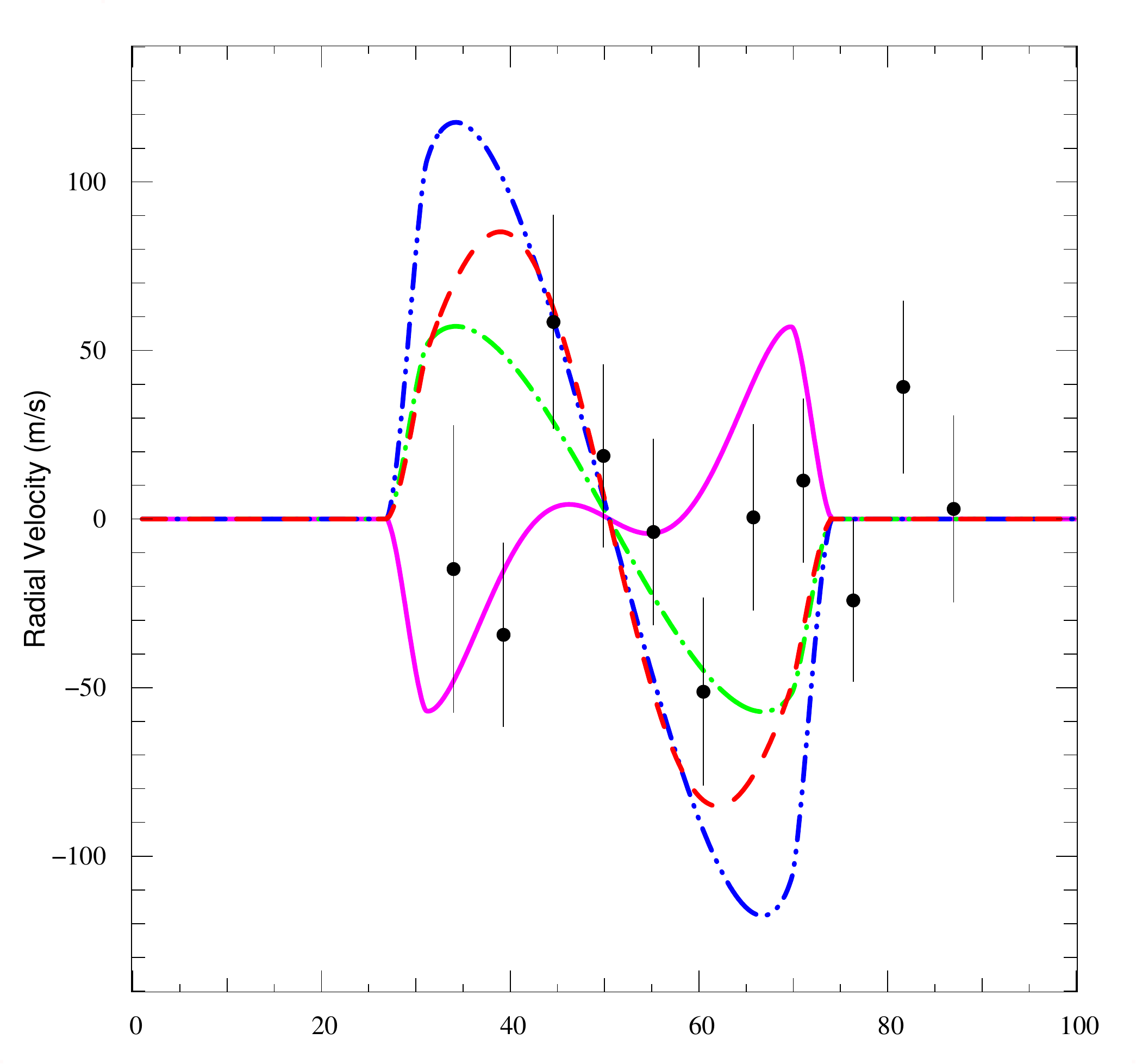}
\caption{{\bf{top}}: For HD\,189733: \textit{dash (red)} is the simulated spectroscopic data with $V\,sin(I) = 3.05\,km.s^{-1}$; \textit{dash double dot (blue)} is a RM model with $V\,sin(I) = 3.30\,km.s^{-1}$; \textit{dash dot (green)} is the same model for a $V\,sin(I) = 3.05\,km.s^{-1}$; \textit{solid (magenta)} is representing the residuals expected by subtracting the RM model by the simulated data; in \textit{(black) points}, a sequence of residuals from Fig.~\ref{fig:HD189} with \textit{Harps}. These residuals are reproduced!
{\bf{bottom}}: For CoRoT-3b: \textit{dash (red)} is the simulated spectroscopic data with $V\,sin(I) = 17\,km.s^{-1}$; \textit{dash double dot (blue)} is a RM model with $V\,sin(I) = 35\,km.s^{-1}$; \textit{dash dot (green)} is the same model for a $V\,sin(I) = 17\,km.s^{-1}$; \textit{solid (magenta)} is representing the residuals expected by subtracting the RM model by the simulated data; with the \textit{(black) points}, a sequence of residuals from Fig.~\ref{fig:corot3} with \textit{Harps}. 
}\label{fig:b}

\end{figure}

We created a grid on a star. Each element of the grid had two pieces of information on it: its intensity (taking limb darkening into account) and its apparent velocity on the line of sight. For the theoretical model, a planet was passed with similar characteristics as the two planets studied in this paper and the centroid of velocity weighted by the light was found for each position thereby recreating the RM effect. For the simulated data, a planet was also passed and a CCF of the star minus the contribution hidden by the planet was generated for each position assuming a Gaussian spectral line for each point on the star. A Gaussian function was fitted to that CCF and its minimum was taken as the simulated RV measurement.

A comparison of the simulated data and the theoretical model showed what W05 had demonstrated: that the simulated data, using the same parameters, has a higher amplitude than the model predicts (see Fig.~\ref{fig:b}). Now, we want to see what a fit of that data would create. All transit parameters are heavily constrained by the photometry save the spin-orbit angle and the $V\,sin(I)$. Assuming a $\beta$ of zero we determined what would happen if only $V\,sin(I)$ was varying. 

On HD\,189733, our test-case, by changing the $V\,sin(I)$ value of the simulated data and comparing it to a model using the found value of $3.30\,km.s^{-1}$ (see Table~\ref{tab:params}), we achieved a very good agreement when $V\,sin(I) =  3.05\,km.s^{-1}$. By subtracting one from the other, we obtained the theoretical residuals expected for a fit such as that produced by the MCMC. The observed residuals from Fig.~\ref{fig:HD189} were added on the graph (Fig.~\ref{fig:b}) and showed a match, therefore indicating that a likely reason was found for these residuals.

The error bars presented in Table~\ref{tab:params} seem to exclude such a large difference on $V\,sin(I)$ at first sight, but it must be remembered that error bars depend on the model used; if the \textit{a priori} is wrong, then the \textit{a posteriori} must be too. The presence of similar residuals in Fig. 1 of W06 shows that not everything is corrected with their method. Nevertheless, the value of $V\,sin(I)$ ($2.97\pm 0.20\,km.s^{-1}$) is similar to what this study infers showing the method developed by W06 is a good tool to estimate $V\,sin(I)$.

The data for HD\,189733b were good enough to allow an adjustment. In the case of CoRoT-3b, where the sizes of error bars are large, such an adjustment was not possible. Instead we compared a theoretical model with the $V\,sin(I)$ value found in Table~\ref{tab:params} to simulated data with the value found by spectral analysis: $17\pm1\,km.s^{-1}$ (D08). The expected residuals and the residuals of the \textit{Harps} sequence were added on top (see Fig.~\ref{fig:b}). Were it not for the abnormally large value of $V\,sin(I)$  found by the fit, which gave clues that something might be going wrong, little information would have been extracted to support the idea the model is not adapted to the data because of error bars of the same order of magnitude as the RM effect itself, let alone the residuals! The absence of a clear signal due to a misfit of the model in the residuals is not enough to rule out the conclusions found with the case of HD\,189733b that indeed the correlated residuals are caused by the imposition of a model not adapted to the data, or vice-versa that the data is not extracted with the same assumptions taken in the model.

Two main paths are now being investigated to rectify the discovered problem: altering the models to take the effect into account, which is the path already taken by W05 (but incomplete as the same residuals are observed), or finding alternative ways of looking at the data to extract it properly. It may be worth noting than tackling asymmetries in the CCF is underway in various areas, be it to understand stellar spots or classical  cepheids stars \citep{Nardetto:2006p1508}. The RM effect is gaining popularity as a new powerful tool to estimate the rotation of stars. Potentially, on misaligned objects like XO-3b \citep{Hebrard:2008p226,Winn:2009p1701} HD\,80606b \citep{Moutou:2009p1081} and CoRoT-3b, if confirmed, we could study differential rotation of stars other than the Sun. Before this can be achieved, however, we have to make sure that the RM effect is properly fitted.

\section{Discussion and Conclusions}

The results that we find are in accordance with previously published results, $V\,sin(I)$s aside. 

A spin-orbit angle is obtained for an exoplanet with unprecedented precision with the value $\beta = 0.85 ^{\circ\,+ 0.32}_{\,\,\,\,- 0.28}$ and is a $3\,\sigma$ detection of an angle different from zero for HD\,189733b's orbit. We confirm a marginal eccentricity which has been strengthened by the addition of the \textit{Harps} data. It remains consistent with values found in \citet{Agol:2008p476} and B09.  Both values are important to constrain planetary formation and evolution models. Results are consistent with the star having no radial velocity drift with time with $\Gamma = 0.2^{+ 2.7}_{- 3.9}\,m.s^{-1}.yr^{-1}$. A comparison with a fit fixing $\Gamma$ to zero showed that the variation in $\chi^2$ was not significant: $\Delta \chi^2 = 10$ for an error on $\chi^2$ of 30. The $V\,sin(I)$ value found by the fit is spurious, as demonstrated in section~\ref{subsec:model}. The real value is probably closer to $3.05\,km.s^{-1}$, value found for the simulated data to explain the theoretical models.

On CoRoT-3b, the fit yields an implausibly high $V\,sin(I)$ at $35\pm8\,km.s^{-1}$, a two fold increase compared to the spectral analysis value, which is only explainable because of a discrepancy of assumptions between the model and the extraction of the data. This overestimation of $V\,sin(I)$ can also be read in the literature: in \citet{Loeillet:2008p881} a $V\,sin(I)$ of $29.5\pm3\,km.s^{-1}$ is found for Hat-P-2 by fitting the RM effect without a prior. This value has to be compared to the value of $21\pm1\,km.s^{-1}$ found with an analysis of the spectral lines and of the photometry. Similarly for CoRoT-2b, in \citet{Bouchy:2008p229}, a $V\,sin(I)$ extracted from the RM model is found to be larger than its spectral and photometric analyses counterparts. The discrepancy between the data and the model has for the moment mostly been observed in the case of fast stellar rotators. Yet, this effect  is also dependent on the $R_p/R_{\star}$ ratio. This means that in the case where the RM effect is fitted independently of photometry leaving every parameter free, this mismatch between the centre of a fitted gaussian and the true velocity of the star's light centroid could lead to incorrect fitting of RM effects on slowly-rotating host stars of planets with high $R_p/R_{\star}$ ratios, and be mistaken for a $V\,sin(I)$ problem. 

On CoRoT-3b we also get a marginal 97\,\% probability detection (equivalent to $2\,\sigma$) of an asymmetry in the current RM data. This result, if confirmed could shed a light into the debated origins and the uncertain nature of CoRoT-3b, a Brown Dwarf according to its mass and the presence of Deuterium burning, but a planet if created by accretion. A tormented history, reflected by a misaligned orbit, could point towards a more planetary origin for this body, notably via coalescence of large planetary bodies as outlined in \citet{Baraffe:2008p1608}, who also point out the confusion of having Deuterium burning as the only limit between Brown Dwarfs and planets. The measure of $\beta$ could become a way of segregating Brown Dwarfs from planets.

Our study of HD\,189733b began as a test-case for combining  two different data sets. It also tested the RM model. CoRoT-3b came as an application of the analysis developed for HD\,189733b and validates that a problem exists between the current models and the way the RV data is extracted.

This paper shows the significance of fitting the various data sets in a combined way. In conclusion this combined analysis, (1) helps reduce the number of free parameters applied to fitting all the data, (2) breaks correlations between some parameters as they are fitted differently, hence exploring parameter space better and (3) insures a consistency on the transit parameters between the models and helps reveal model inconsistencies, as it did here with the case of the Rossiter-McLaughlin effect, therefore ensuring that systematics are not mistaken for physics.

\begin{acknowledgements} 
The authors would like to acknowledge the use of the ESO archive, ADS, \textit{Simbad} and \textit{Vizier} but mostly the help and the kind attention of the ESO staff at La Silla. The work is supported by the Swiss Fond National de Recherche. A. Triaud would like to acknowledge that he owes a beer to D. Queloz and  two to A. Collier Cameron. We also thank the referee for the useful suggestions provided.
\end{acknowledgements} 

\bibliographystyle{aa}
\bibliography{11897txt}

\end{document}